\documentclass[10pt,journal,compsoc]{IEEEtran}

\usepackage{ifpdf}

\ifCLASSOPTIONcompsoc
  \usepackage[nocompress]{cite}
\else
  \usepackage{cite}
\fi
\usepackage{microtype}                 
\PassOptionsToPackage{warn}{textcomp}  
\usepackage{textcomp}                  
\usepackage{tabu}                      
\usepackage{booktabs}                  
\usepackage{enumitem}
\usepackage{url}
\usepackage{stfloats}
\usepackage{multirow}
\usepackage{booktabs}
\usepackage{xcolor}
\usepackage[xcolor=dvipdf]{changes}
\usepackage{ulem}
\usepackage{wrapfig}

\definecolor{relatedorange}{HTML}{E07734}
\definecolor{lightblue}{HTML}{EAEEFF}
\definecolor{lightyellow}{HTML}{FEEBAD}
\definecolor{deletegray}{HTML}{B8B9BA}

\PassOptionsToPackage{svgnames,x11names,dvipsnames}{xcolor}
\usepackage[most]{tcolorbox}
\newtcbox{\inlineboxTop}[1][]{enhanced,
 box align=base,
 nobeforeafter,
 colback=white,
 colframe=relatedorange,
 sharp corners,
 size=small,
 left=2pt,
 right=2pt,
 top=3pt,
 bottom=0pt,
 boxsep=0pt,
 #1}

\newtcbox{\inlineboxBottom}[1][]{enhanced,
 box align=base,
 nobeforeafter,
 colback=white,
 colframe=relatedorange,
 sharp corners,
 size=small,
 left=2pt,
 right=2pt,
 top=0.5pt,
 bottom=2.5pt,
 boxsep=0pt,
 #1}

\newtcbox{\inlineboxall}[1][]{enhanced,
 box align=base,
 nobeforeafter,
 colback=white,
 colframe=relatedorange,
 sharp corners,
 size=small,
 left=2pt,
 right=2pt,
 top=0.5pt,
 bottom=0.5pt,
 boxsep=0pt,
 #1}

\setdeletedmarkup{\color{gray}{\sout{#1}}}
\setaddedmarkup{\color{orange}{#1}}

\newcommand{\NLI}{{NLI}}
\newcommand{\name}{{XNLI}}
\newcommand{\ie}{\emph{i.e.,}}
\newcommand{\eg}{\emph{e.g.,}}
\newcommand{\icon}[1]{\includegraphics[height=\fontcharht\font`\B]{#1}}

\newcommand{\feng}[1]{{\color{black} #1}}
\newcommand{\gram}[1]{{\color{black} #1}}

\ifCLASSINFOpdf
\else
\fi
\newcommand\MYhyperrefoptions{bookmarks=true,bookmarksnumbered=true,
pdfpagemode={UseOutlines},plainpages=false,pdfpagelabels=true,
colorlinks=true,linkcolor={black},citecolor={black},urlcolor={black},
pdftitle={Bare Demo of IEEEtran.cls for Computer Society Journals},
pdfsubject={Typesetting},
pdfauthor={Michael D. Shell},
pdfkeywords={Computer Society, IEEEtran, journal, LaTeX, paper,
             template}}
\ifCLASSINFOpdf
\usepackage[\MYhyperrefoptions,pdftex]{hyperref}
\else
\usepackage[\MYhyperrefoptions,breaklinks=true,dvips]{hyperref}
\usepackage{breakurl}
\fi
\begin{document}
%
\title{XNLI: Explaining and Diagnosing NLI-based Visual Data Analysis}
%
%
%
%

\author{Yingchaojie Feng,
        Xingbo Wang,
        Bo Pan,
        Kam~Kwai~Wong,
        Yi Ren,
        Shi Liu,
        Zihan Yan,\\
        Yuxin Ma,
        Huamin Qu,
        and Wei Chen.
\IEEEcompsocitemizethanks{
\IEEEcompsocthanksitem Y. Feng, B. Pan, Y. Ren, S. Liu, W. Chen are with The State Key Lab of CAD \& CG, Zhejiang University, Hangzhou, Zhejiang, China.
W. Chen is also with the Laboratory of Art and Archaeology Image (Zhejiang University), Ministry of Education, China.
W. Chen is the corresponding author.
E-mail: \{fycj, bopan, rymaster, zju\underline{~~}ls, chenvis\}@zju.edu.cn.
\IEEEcompsocthanksitem X. Wang, KK Wong, H. Qu are with the Hong Kong University of Science and Technology, Hong Kong, China.
Email: \{xwangeg, kkwongar, huamin\}@cse.ust.hk.
\IEEEcompsocthanksitem Z. Yan is with the MIT Media Lab, Cambridge, MA, USA.
Email: yzihan@media.mit.edu.
\IEEEcompsocthanksitem Y. Ma is with the Department of Computer Science and Engineering, Southern University of Science and Technology, Guangdong, China.
Email: mayx@sustech.edu.cn.
}
\thanks{Manuscript received April 19, 2005; revised August 26, 2015.}}

%
%

\markboth{IEEE TRANSACTIONS ON VISUALIZATION AND COMPUTER GRAPHICS}%
{Shell \MakeLowercase{\textit{et al.}}: Bare Advanced Demo of IEEEtran.cls for IEEE Computer Society Journals}
%



\IEEEtitleabstractindextext{%
\begin{abstract}
Natural language interfaces (NLIs) \gram{enable users to flexibly specify} analytical intentions in data visualization. However, \gram{diagnosing} the visualization results without understanding the underlying generation process \gram{is challenging}. Our research explores how to provide explanations for {\NLI}s to help users locate the problems and further revise the queries. We present XNLI, an explainable NLI system for visual data analysis. The system introduces a Provenance Generator to reveal the detailed process of visual transformations, a suite of interactive widgets to support error adjustments, and a Hint Generator to provide query revision hints \gram{based on} the analysis of user queries and interactions. Two \gram{usage scenarios of {\name}} and a user study verify the effectiveness and usability of the system. Results suggest that XNLI can significantly enhance task accuracy without interrupting the NLI-based analysis process.
\end{abstract}

}

\maketitle


\IEEEdisplaynontitleabstractindextext

%
\IEEEpeerreviewmaketitle

\ifCLASSOPTIONcompsoc
\IEEEraisesectionheading{\section{Introduction}\label{sec:introduction}}
\else
\section{Introduction}
\label{sec:introduction}
\fi

\IEEEPARstart{N}{atural} Language Interfaces ({\NLI}s) \gram{have received} widespread attention in academia \cite{shen2021towards, setlur2016eviza, hoque2017applying, gao2015datatone, narechania2020nl4dv} and industry \cite{tory2019mean, powerbi}.
\gram{In} visual data analysis, typical {\NLI}s \cite{shen2021towards, narechania2020nl4dv, gao2015datatone} allow users to pose natural language (NL) queries to specify visualizations \gram{from} data attributes, \gram{analytical} tasks, and visual encodings.
To better meet user expectations, previous studies have enhanced the query interpretation capability of {\NLI}s, such as parsing long compound statements~\cite{hoque2017applying}, resolving ambiguities~\cite{gao2015datatone}, inferring underspecification~\cite{setlur2019inferencing}, and understanding contextual references~\cite{hoque2017applying}. 

However, \gram{as} the visualization generation process contains multi-step data transformations and visual encodings \cite{card1999readings}, \gram{users have difficulty diagnosing} the final visualization results without understanding the underlying generation process.
Considering a movie exploration scenario, a user poses a query to select the movie records within a rating range and then visualize their average budgets \gram{across} genres in a bar chart.
\gram{As} the rating values are not encoded in the resulting chart, the user cannot assess the correctness of the filtering task.
In addition, the lack of knowledge about the {\NLI} impedes users from correcting unexpected results.
Thus, users may fail to derive desired insights from data and lose trust in {\NLI} tools.
Both of these issues hinder the process of {\NLI}-based data analysis and decision-making. 

\feng{Prior studies \cite{gao2015datatone, setlur2016eviza, hoque2017applying, srinivasan2020interweaving} have introduced interactive widgets to help users identify ambiguities in the queries and support adjustments.
Nevertheless, \gram{diagnosing} the unexpected results caused by failures of query interpretation in the {\NLI} pipeline or impractical visualizations specified in the queries \gram{remains a challenge.}
\gram{In this study, we} investigate the major causes of the unexpected results and summarize them with the key aspects of the {\NLI} process ({\ie} attributes, tasks, and visual encodings).}
To help users understand the {\NLI} process, identify problems, and possibly fix them during data analysis, we formulate a set of design requirements and develop {\name}, an explainable {\NLI} that elucidates the visualization generation process, supports interactive adjustments, and provides hints for query revision.

\gram{In addition to} interpreting user queries and generating visualization results, the system reveals the process from the aspects of the attributes, tasks, and visual encodings and leverages data provenance to facilitate user understanding.
The relevant trigger words in the original queries and the inference uncertainty during the query interpretation are visualized \gram{for examining the} potential errors in the process and \gram{locating their origins}.
Based on \gram{this information}, the system provides interactive widgets, allowing users to intervene and adjust the visualization process.
To facilitate the use of {\NLI} for the desired data analysis, the system helps users refine their queries by providing two types of hints.
Rule-based hints point out the potential problems in the queries regarding attributes, tasks, and visual encodings.
Interaction-based hints are generated based on the analysis of users' interactive adjustments, and contain query revision suggestions as well as valid query examples that can be \gram{applied} in the subsequent query formulation.

Finally, we introduce two \gram{usage scenarios} to illustrate how our system helps users identify and correct different kinds of problems in the process, such as unexpected attributes and tasks, and gain knowledge about the NLI capability (e.g., how to phrase system-supported queries).
We also conduct a controlled user study with a baseline to evaluate the effectiveness and usability of our system and the impact of explanations on the analysis process. The study results suggest that our system is easy to use, and the explanations effectively improve the \gram{analytical} efficiency without interrupting the analysis process.
In summary, our contributions include:
\begin{itemize}[noitemsep, topsep=0pt]
    \item A pilot study for {\NLI}-based visual analysis that identifies challenges in process understanding and summarizes the major problems from the {\NLI} aspects.
    \item An {\NLI}-based visual analysis system that explains the {\NLI} process, supports interactive error correction, and provides hints to help query formulation and revision. 
    \item Two \gram{usage scenarios} and a controlled user study to evaluate the effectiveness and usability of our system.
\end{itemize}

\section{Related Work}
\subsection{NLI for Data Visualization}
NLIs are widely studied as a promising means of interaction for visual analytics \cite{srinivasan2017natural, shen2021towards}.
NLIs for data visualization generate visual representations in response to NL queries, which help reveal data insights \cite{yu2019flowsense, srinivasan2017orko, setlur2020sneak, setlur2021geosneakpique, chen2022crossdata, chen2022sporthesia, 9906966, chen2022nebula}.
Cox et al. \cite{cox2001multi} \gram{aimed} to integrate {\NLI} into existing visualization systems. 
Subsequently, \gram{extensive} research has focused on understanding complex and free-form NL queries, such as incomplete or referential utterances \cite{setlur2016eviza, hoque2017applying} and conversational queries \cite{setlur2016eviza, hoque2017applying, kumar2016towards, fast2018iris}.
NL4DV \cite{narechania2020nl4dv} is a general {\NLI} toolkit that integrates a set of NLP techniques ({\eg} dependency parsing) to interpret NL data queries and recommends visualizations to generate results.
Snowy \cite{srinivasan2021snowy} and QRec-NLI \cite{wang2022interactive} recommend utterances in NL-based visual analysis, leveraging NL to provide analytical guidance.
\feng{In recent years, {\NLI}s have been applied \gram{to} various scenarios and tasks, such as flow data exploration \cite{huang2022flownl}, visualization authoring \cite{wang2022towards}, and comparative analysis \cite{gaba2022comparison}.}
\gram{Other studies} \cite{fu2020quda, luo2021synthesizing, srinivasan2021collecting, luo2021nvbench} build query corpora for linguistic \gram{property} analysis and construct labeled benchmark datasets for {\NLI} evaluation and deep learning model training \cite{luo2021natural, liu2021advisor}.

In addition to enhancing the capability of query interpretation, \gram{revealing} the {\NLI} process and enabling feedback for error correction \gram{are also necessary}.
DataTone \cite{gao2015datatone} introduces ambiguity widgets highlighting ambiguities in query interpretation and supporting interactive specification.
Such interactive widgets have been widely integrated into later works \cite{setlur2016eviza, hoque2017applying, srinivasan2017orko, srinivasan2020interweaving}.
Commercial {\NLI} systems, such as Power BI \cite{powerbi} and Tableau \cite{tory2019mean}, present the intermediate parsing results as structured text and enable interactive adjustments of the chart results.
Our study provides systematic process explanations from the key aspects of the general {\NLI} pipeline \cite{shen2021towards, narechania2020nl4dv} and demonstrates data provenance to help users understand the process. 
Besides supporting widget-based adjustments, our system provides suggestions for query rephrasing and revision, facilitating the use of {\NLI} in the subsequent analysis process.

\subsection{Provenance}
Provenance records the evolution of the analysis process and has been widely studied in the visualization field. Ragan et al. \cite{ragan2015characterizing} summarize an organizational framework for the provenance of information (data, visualization, and interaction) and purposes (collaborative communication, presentation, and meta-analysis).
Heer et al. \cite{heer2008graphical} abstract the provenance as a node-link diagram and propose a graphical history tool with provenance to support analysis, communication, and evaluation.
Provenance can be used in data wrangling \cite{bors2019capturing, kasica2020table, xiong2022visualizing} and analytical reasoning \cite{shrinivasan2008supporting}.
Datamations \cite{pu2021datamations} maps the data operations in the provenance to \gram{various} types of animations and demonstrates the benefits of data animation \gram{to explain} the Simpson's paradox.

Some works employ provenance to explain complex codes.
QueryVis \cite{leventidis2020queryvis} generates logic-based diagrams to represent the meaning of deeply nested SQL statements.
Berant et al. \cite{berant2019explaining} design a cell-based provenance with \gram{NL} explanations to accelerate user understanding of complex queries.
Data Tweening \cite{khan2017data} introduces a grammar to generate incremental visualizations for queries of data transformation.
A recent work, DIY \cite{narechania2021diy}, leverages the representative sample data to demonstrate the transformation process during the execution of SQL statements.

We base our work on provenance to facilitate user understanding of the {\NLI} process and support interactive adjustments. Moreover, we highlight the inference uncertainties and ambiguities during the query interpretation to guide users in investigating the potential problems in the process, further improving the efficiency of problem identification.

\section{Design Process}
The target users of our system are data analysts who leverage {\NLI}s to perform visual analysis.
The design process lasted 9 months and started with a pilot study to investigate the challenges of understanding and diagnosing visualization results in {\NLI}-based analysis scenarios.
We summarized the unexpected system results in the open-ended task exploration and categorized them into the major cause types from the perspective of the {\NLI} process.
We iteratively refined the design requirements based on user feedback and built corresponding system prototypes for verification.

\subsection{Experiment Design}
\textbf{Participants.}
We recruited 9 data analysts (two females and seven males) from a local university. 
Seven participants have experience in visual analysis, with one having prior experience in using {\NLI}. The remaining two are beginners in visual analysis.

\textbf{Settings.}
We developed an {\NLI} prototype system for visual analysis.
It is powered by NL4DV \cite{narechania2020nl4dv}, a typical {\NLI} toolkit for tabular data analysis.
The system enables users to preview the dataset and pose NL queries.
We used the IMDb movie dataset, which contains 710 movie items with nine attributes. 
Participants were requested to perform open-ended task exploration \cite{srinivasan2020ask} over the dataset to provide investment advice. 

\textbf{Procedure.}
We first introduced the study settings and workflow of the system prototype.
After the participants familiarized themselves with the system, they performed the open-ended task exploration.
We provided minimal support during the sessions and let participants obtain their desired \gram{analytical} results through trial and error.
Then, we conducted a post-study interview with participants to reflect on the exploration experience.

\subsection{Results Analysis}
\label{sec:pilot}
During the post-study interview, we explained to the participants the provenance of the visualization results obtained during the exploration process, from which we identified some system results that did not match the user's expectations.
For these cases, we asked \gram{the} participants to revise their queries to get the desired visualizations with a time constraint.
Finally, following the bottom-up approach in the prior study, \cite{tory2019mean, wambsganss2020adaptive, xia2022persua, wang2021dehumor}, we summarized the major causes of the unexpected system results from the aspects of attributes, tasks, and visual encodings. \gram{Furthermore}, we analyzed the participants' ability to locate and correct unexpected system results.
The results are shown in \autoref{tab:unexpected}.

\begin{table}[ht]
\setlength{\abovecaptionskip}{0cm}
\caption{Major types of unexpected results that we summarized from three key aspects ({\ie} attributes, tasks, visual encodings) of the {\NLI} workflow.}
\label{tab:unexpected}
\resizebox{\linewidth}{!}{
\renewcommand{\arraystretch}{1.5}
\begin{tabular}{lllll}
\hline
\textbf{Aspects} & \textbf{Unexpected Behaviors} & \textbf{Total} & \textbf{Locate} & \textbf{Correct} \\ \hline
\multirow{2}{*}{Attributes}        & Fail to infer             & 17    & 12    & 9               \\ \cline{2-5}
                                            & Unexpected types          & 14    & 5     & 5               \\ \hline
\multirow{3}{*}{Tasks}             & Fail to infer             & 9     & 6     & 1               \\ \cline{2-5}
                                            & Unexpected types          & 15    & 9     & 4               \\ \cline{2-5}
                                            & Impractical parameters    & 12    & 5     & 4               \\ \hline
\multirow{2}{*}{Visual Encodings}  & Chart types               & 8     & 6     & 4               \\ \cline{2-5}
                                            & Encoding channels          & 5     & 5     & 3               \\ 
\hline
\end{tabular}}
\end{table}

For \textbf{attributes}, {\NLI} usually \textit{failed to infer} the attributes.
Participants inputted typos or used synonyms, and NLI failed to match them with the corresponding attributes.
Fortunately, these issues could usually be identified due to the absence of attributes in the resulting charts and were easy to correct. 
\textit{Unexpected types} of attributes were typically caused by incorrect implicit inference.
For example, the vague modifier ``high'' in the query was treated by the {\NLI} as the specific item of the ``Title'' attribute, 
which also led to an unexpected filtering ({\ie} only choosing the movies with ``high'' in the title). 
Participants felt something wrong but always could not locate the problems or correct them.

For \textbf{tasks}, NLI might \textit{fail to infer} their types, including unsupported tasks ({\eg} finding the extremes) and task-related keywords ({\eg} failing to understand ``at least''). 
\textit{Unexpected types} of tasks occurred when multiple task schemas \gram{were available}. A participant posed, ``show the correlation of gross and IMDB rating'', expecting the average gross of the binned IMDB rating in a bar chart instead of all data items in a scatterplot.
Participants sometimes specified tasks with \textit{impractical parameters} ({\eg} out-of-range filter conditions) due to a lack of knowledge of the data.
As a result, no data records matched the criteria, and empty charts were returned, which always confused the users when trying to locate the problems.
Another reason for \textit{impractical parameters} was the incorrect mappings of tasks and parameters constructed based on  the parsing dependency of NL queries. 
\gram{Participants had difficulty noticing the issues if} the task-related attributes were not encoded in the visualization results.

For \textbf{visual encodings}, unexpected \textit{chart types} made up the majority.
Participants might specify an unsupported chart type and get nothing back. They could locate the problem and expected the system to suggest some available chart types. Unexpected \textit{chart types} also occurred when they were underspecified ({\eg} participants expected a pie chart to visualize the data proportion instead of the provided bar chart.) A few cases of unexpected \textit{encoding channels} were observed ({\eg} participants preferred encoding a quantitative attribute by color instead of size in the scatter plot).

\feng{In summary, the user challenge of process understanding is twofold.
First, it can be difficult to locate the problems of the unexpected results, especially for the underlying attributes and tasks which are not encoded in the final chart results.
Second, it would be time-consuming for inexperienced users to correct the errors and \gram{obtain} the desired results through query revision.
Due to their lack of knowledge about the {\NLI}'s mechanism and capability, users might fail to revise the queries with the correct trigger words and phrases that are well supported by the {\NLI} system.
}

\subsection{Design Requirements}
To help users understand the visualization process and address the \gram{aforementioned} issues, we conducted regular bi-weekly meetings with all participants (\textit{P1}-\textit{P9}) from the pilot study and refined our prototype system iteratively. We also conducted a literature survey \gram{of} the prior work on {\NLI}. Finally, we summarized a set of design requirements.

\textbf{R1: Provide an overview of the visualization process from the key aspects of the {\NLI} pipeline.}
The typical {\NLI}s interpret the user queries and formulate the visualization responses based on three key aspects ({\ie} attributes, tasks, and visual encodings).
As observed in the pilot study, it is challenging for users to identify unexpected results without knowing the {\NLI} pipeline.
Most participants stated that the system should present these key aspects as an overview of the visualization process so that they \gram{could} go through the process and dive into some aspects of interest.

\textbf{R2: Leverage the representative data items to demonstrate the visualization process.} 
\textit{P2} pointed out that explaining the process with data \gram{could} promote understanding and increase trust.
However, presenting all the contents of a large dataset \gram{causes} serious visual clutter beyond human recognition affordance \cite{narechania2021diy, xiong2022visualizing}.
The system should choose the representative data items from relevant data columns and use them to demonstrate the visualization process.
Moreover, visual cues ({\eg} highlighting) are also necessary to guide the user's attention to the differences between steps.

\textbf{R3: Reveal the connections between the visualization process and the original query.} 
Each aspect of the visualization process is specified according to the original query. These potential connections constructed by {\NLI} \cite{chen2022crossdata} are usually inaccessible to the users. 
\textit{P1} and \textit{P5} commented that they wondered how their queries \gram{were} parsed by the system, especially when \gram{obtaining} unexpected results.
The system should link the visualization process to the related parts of the query to help users to investigate the causes of unexpected results and gain inspiration for query revision.

\textbf{R4: Show the inference types of {\NLI} to highlight the uncertainty of query interpretation.} 
{\NLI} employed multiple strategies for query interpretation \cite{dhamdhere2017analyza, gao2015datatone, yu2019flowsense, srinivasan2020interweaving} \gram{because of} the diversity of NL expression.
For example, data attributes may be explicitly or implicitly mentioned in the query, which require \gram{various} approaches for inference \cite{shen2021towards} and thus introduces different levels of uncertainty \cite{narechania2020nl4dv}. 
Presenting the inference types and ambiguities makes users aware of the uncertainty in the process and prioritizes attention for further inspection.

\textbf{R5: Support adjustment of the visualization process through directed manipulation.} 
A considerable proportion of cases from the pilot study show that even when users locate the problems of unexpected results, they still face challenges in query revision for desired results.
Directed manipulation can serve as a complementary interaction modality \cite{srinivasan2017orko, saktheeswaran2020touch} to support the further expression of user expectations.
Prior works \cite{gao2015datatone, setlur2016eviza, hoque2017applying, srinivasan2017orko} have shown the great potential of interactive widgets in resolving ambiguity.
The system should provide interactive widgets to support user adjustments to the visualization process. 

\textbf{R6: Provide hints and query examples for query phrasing and revision.} 
Compared to interactive adjustments of current query results, providing hints for query revision has a more positive \gram{effect} on the subsequent iteration of NL interaction. 
\gram{This requirement} leverages unexpected cases as teaching opportunities and \gram{spares} users from the \gram{tedious task} of guessing supported utterances.
\textit{P6} suggested that the system should inform the users about potential errors in the query proactively.
\textit{P1} stated that providing smart and effective suggestions and valid query examples would be better to help him revise the original queries.

\section{System Design}
To meet the derived requirements, we develop {\name}, a prototype system that provides visual explanations for NLI results to facilitate understanding of the visualization generation process and gain knowledge of query phrasing.
In this section, we first describe an overview of the system workflow. Then, we introduce the visual and functional designs of the user interface to demonstrate the usage of explanations for {\NLI}-based visual analysis.
Finally, we \gram{proceed to} the detailed implementation of major system components.

\begin{figure*}[tb]
 \centering 
 \includegraphics[width=\linewidth]{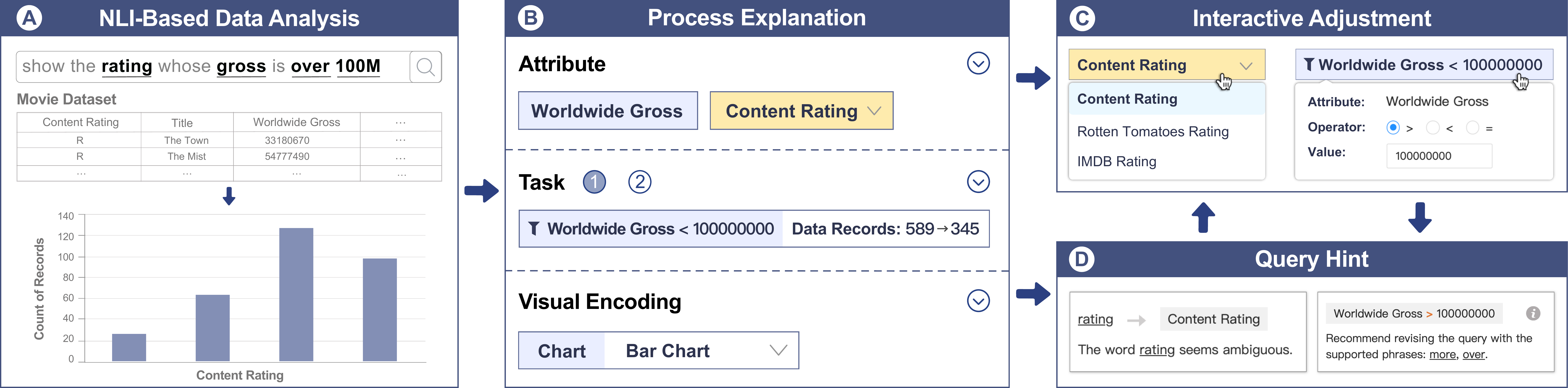}
 \caption{The workflow of our system consists of four components that (A) support {\NLI}-based data analysis, (B) provide process explanations for visualization results from the aspects of attributes, tasks, and visual encodings, (C) \feng{provide interactive widgets to adjust the process} from each aspect, and (D) provide hints to help problem localization and query revision.}
 \label{fig:framework}
 \vspace{-5pt}
\end{figure*}

\subsection{System Workflow}
{\name} is implemented as a web-based system with a graphical user interface to enable easy access.
The system workflow is summarized in \autoref{fig:framework}.
As the basic feature of {\NLI}, our system supports \textbf{NL-based data analysis} (\autoref{fig:framework}$A$).
Users can pose data-related queries through NL and conduct visual analysis based on the generated visualization results.
The query interpretation capability of our system is powered by NL4DV \cite{narechania2020nl4dv}, a general and representative toolkit for {\NLI} construction.
To facilitate user understanding of \gram{the} visualization results, our system provides \textbf{process explanations} (\autoref{fig:framework}$B$) regarding three key aspects of the {\NLI} pipeline, {\ie} attributes, tasks, and visual encodings (R1-R2).
The process explanations build a basis for users to identify unexpected results (R3-R4). The system further provides \textbf{interactive widgets} (\autoref{fig:framework}$C$) for each visualization aspect, leveraging directed manipulation as a complementary interaction modality to support error correction (R5).
To help users \gram{benefit more from} {\NLI}, the system provides \textbf{query hints} (\autoref{fig:framework}$D$) as feedback to indicate potential errors in user queries and suggest valid query phrasing (R6).
The hints are generated based on the analysis of visualization aspects and logs of interactive adjustments.

\begin{figure*}[tb]
 \centering 
 \includegraphics[width=\textwidth]{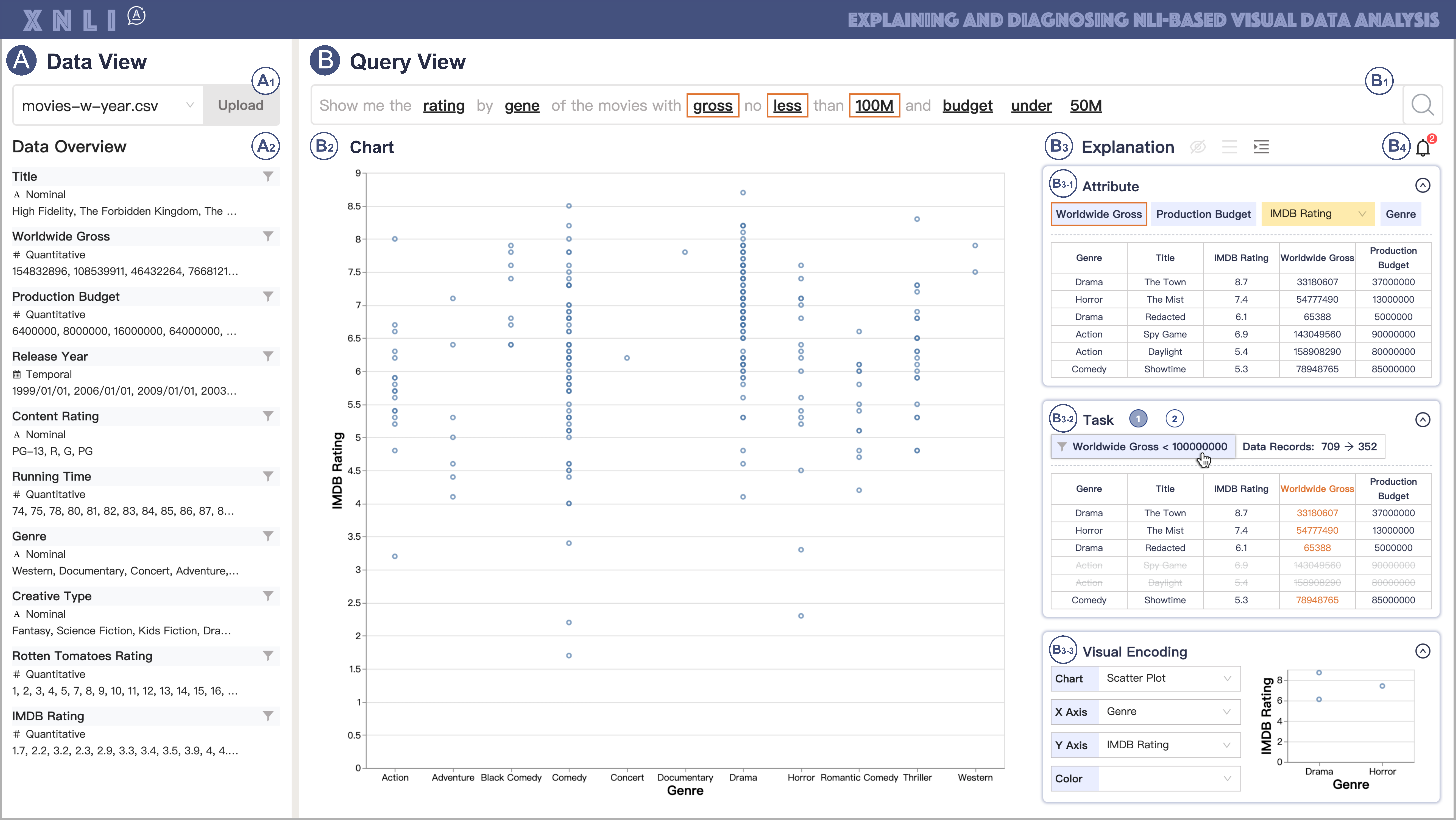}
 \caption{The user interface of {\name} \gram{consists of} two views, \gram{including (A) the Data View and (B) the Query View}. Users can (A1) select \gram{or upload} the dataset, (A2) view the data attributes, (B1) input natural language queries, (B2) use charts for data analysis, (B3) check the explanations of the NLI process and correct the unexpected results through interactive widgets, and (B4) gain hint feedback for query revision.}
 \label{fig:system}
 \vspace{-5pt}
\end{figure*}

\subsection{User Interface}
The user interface (Figure \ref{fig:system}) consists of three modules.
The {\NLI} module allows users to overview the explored dataset (\autoref{fig:system}$A_1$, $A_2$) and pose NL queries for visual data analysis (\autoref{fig:system}$B_1$, $B_2$).
The Explanation module reveals the visual transformation process (\autoref{fig:system}$B_3$) from aspects of the attributes, tasks, and visual encodings.
Each aspect supports interactive adjustments for error correction.
To further help the user derive the desired queries for data analysis, the Hint module provides suggestions for query revision and \gram{pops} up (\autoref{fig:system}$B_4$) \gram{when} the user clicks the icon (\icon{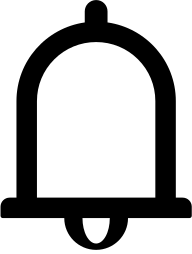}).

\subsubsection{{\NLI} Module}
After users choose or upload a dataset for exploration (\autoref{fig:system}$A_1$), the Data Overview panel presents metadata for each attribute in the dataset (\autoref{fig:system}$A_2$), including the attribute name, type, and typical values.
Attribute types include Quantitative, Nominal, Ordinal, and Temporal \cite{satyanarayan2016vega}. 
\gram{The} typical values \gram{of the attributes} are chosen based on the occurrence rate. 
To the right of each attribute name is a filter icon that users can click to add a filtering task for the attribute to the visualization process (R5).

The input box (\autoref{fig:system}$B_1$) at the top of Query View allows users to type NL queries.
After \gram{the query interpretation}, the keywords in the sentence will be encoded in different styles according to their usage.
The keywords related to attributes, tasks, and visual encoding are encoded in black and underlined, while other words are encoded in grey. 

The Chart panel (\autoref{fig:system}$B_2$) presents the visualization results in response to user queries.
It is the main window for data analysis.
The charts are specified by Vega-Lite and support user interactions, such as hovering and selection.

\subsubsection{Explanation Module}
The Explanation panel (\autoref{fig:system}$B_3$) presents the visualization process from aspects of the attributes, tasks, and visual encodings (R1).
Each aspect supports overview and details-on-demand.
The detailed level can be collapsed (\icon{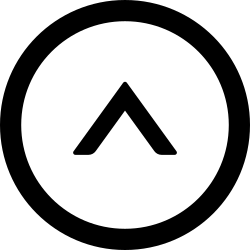}) and opened (\icon{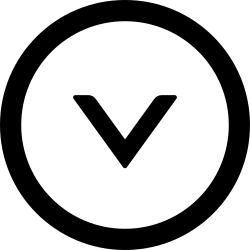}).
Users can choose the default display modes of the explanations in the title bar, including \textit{No Explanation} (\icon{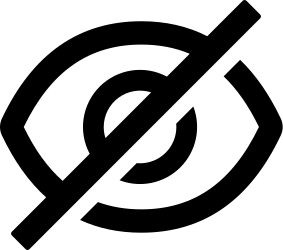}), \textit{Overview Explanation} (\icon{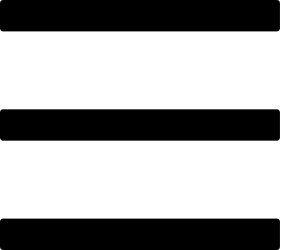}), and \textit{Detailed Explanation} (\icon{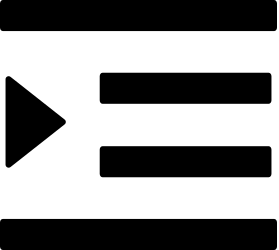}).

The Attribute aspect (\autoref{fig:system}$B_{3-1}$) lists all the attribute names involved and visualizes the inference types in different background colors (R4).
\feng{The attributes that are explicitly inferred ({\eg} Genre in \autoref{fig:system}$B_{3-1}$) are encoded in light blue, while the attributes that are implicitly inferred ({\eg} Title in \autoref{fig:case_one}$A_3$) or ambiguous ({\eg} IMDB Rating in \autoref{fig:system}$B_{3-1}$) are encoded in yellow.}
The ambiguous attributes are augmented with interactive selectors \feng{(\autoref{fig:framework}$C$)} to \gram{support adjustments} (R5).
User preference for ambiguities indicated by explicit adjustments or implicit agreement will be used for subsequent query disambiguation.
The detailed level of the Attribute aspect provides a sample dataset to demonstrate the visualization provenance (R2).

The Task aspect (\autoref{fig:system}$B_{3-2}$) shows the data transformation tasks and the corresponding changes in the actual amount of data.
\gram{For example}, the amount of data is reduced from 709 to 352 during the filtering task.
The background color of the task ``Worldwide Gross $<$ 100,000,000'' visualizes the inference types from queries (R4), just like the attributes.
Hovering on the task triggers the highlighting of the related keywords ``\textit{gross}'', ``\textit{less}'', and ``\textit{100M}'' in the query (R3), \gram{helping} users discover that the unexpected task is caused by a misinterpretation of ``no less''.
Users can click the task (\autoref{fig:framework}$C$) and fix the error in the pop-up box (R5).
The detailed level of the Task aspect leverages visual cues to emphasize the changes of sample data provenance (R2). 
For example, the task is performed on the highlighted attribute Worldwide Gross, and some data items are filtered out.

The Visual Encoding aspect (\autoref{fig:system}$B_{3-3}$) presents chart type and specific encoding channels.
Users can change any mappings through directed manipulation (R5).
The encoding results of sample data (after data transformation tasks) are shown on the right to aid in understanding and validation.

\subsubsection{Hint Module}
\begin{figure}[t]
 \centering 
 \includegraphics[width=\linewidth]{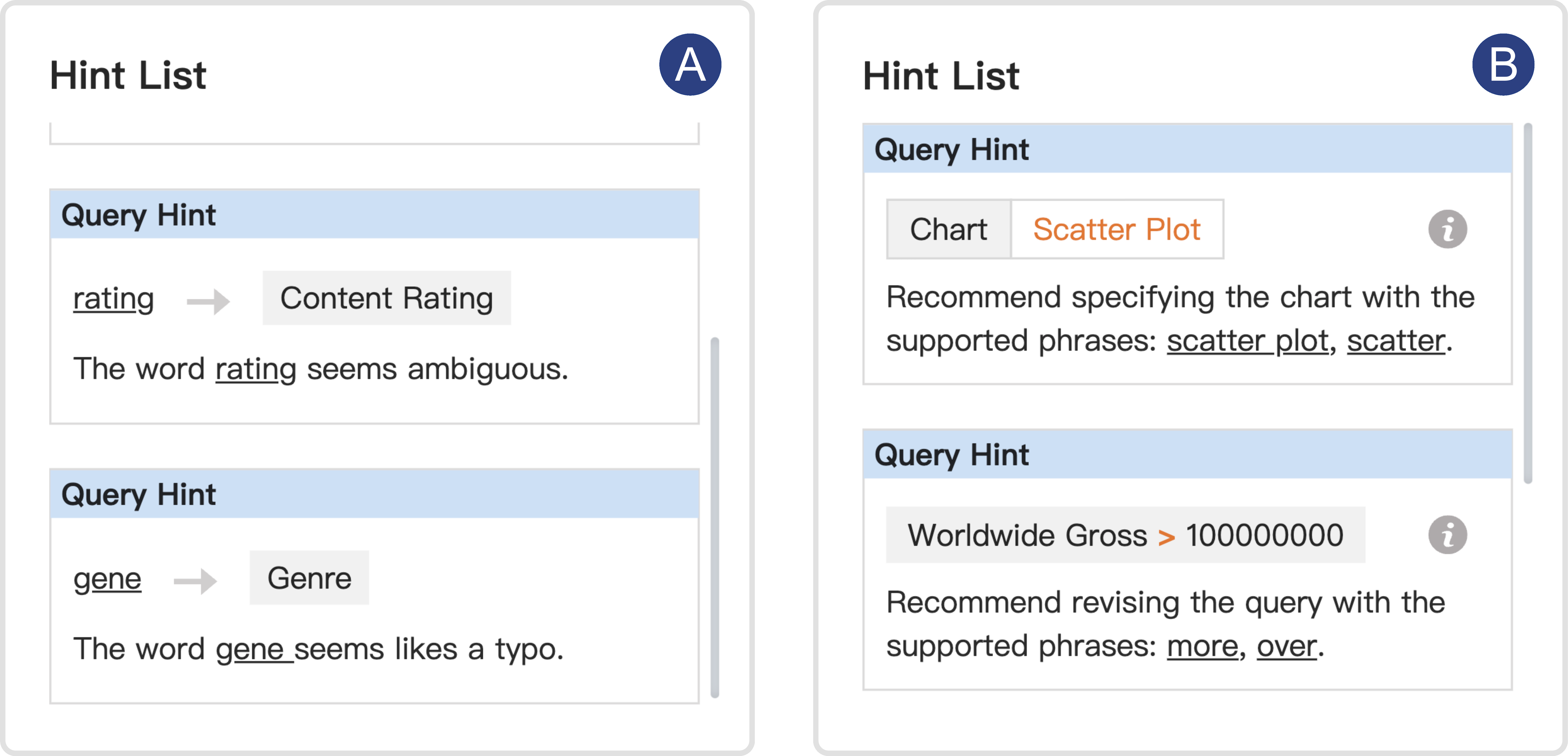}
 \caption{Hint module presents (A) rule-based hints for issues in queries (\eg~input typos and ambiguities) and (B) interaction-based hints after user adjustments through interactive widgets.}
 \label{fig:hint}
 \vspace{-5pt}
\end{figure}

\gram{The} Hint panel (\autoref{fig:hint}) provides two \gram{types} of hints for query revision, {\ie} rule-based hints and interaction-based hints.
Rule-based hints (\autoref{fig:hint}$A$) indicate the potential errors in the queries (R6), such as input typos and ambiguities.
The system detects these errors during the query interpretation and provides the hints alongside the generated visualization results.
Other errors, such as unexpected types of attributes and tasks, are \gram{due to} unexpected query interpretations \gram{that} are unknown to the system.
Therefore, the system provides interaction-based hints (\autoref{fig:hint}$B$) after user adjustments \gram{through} interactive widgets.
The interaction-based hints contain some targeted suggestions for local query modifications, as well as valid query examples that correspond to the adjusted visualization results and can be used in subsequent data analysis (R6).

\subsection{Provenance Generator}
\begin{figure*}[t]
 \centering 
 \includegraphics[width=\linewidth]{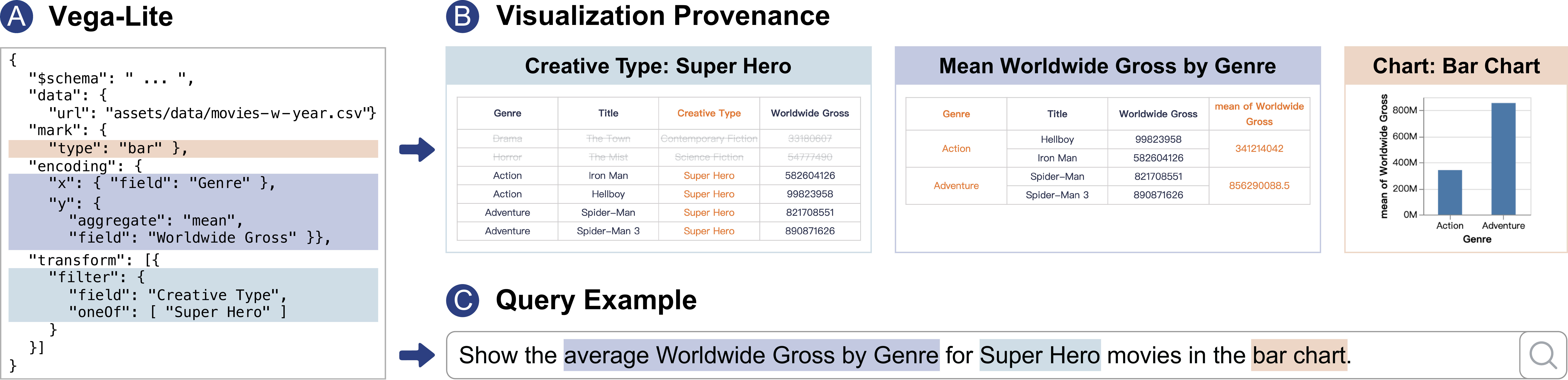}
 \caption{{\name} leverages the (A) Vega-Lite specification to generate (B) visualization provenance to explain the process of visualization generation and (C) query examples as hints \gram{to help users revise the queries}.}
 \label{fig:approach}
 \vspace{-5pt}
\end{figure*}

The visualization results of NL4DV are specified with Vega-Lite grammar (\autoref{fig:approach}$A$), which is popular in the field of {\NLI} \cite{narechania2020nl4dv, wang2022interactive, srinivasan2021snowy, luo2021natural, luo2021synthesizing, srinivasan2021collecting, islam2022natural, fu2020quda}.
We leverage provenance to depict the three aspects of the visualization process and introduce  the Provenance Generator, which \gram{consists of} three major components to construct the visualization provenance from the Vega-Lite grammar, select representative sample data for demonstration, and apply visual cues to highlight the changes \gram{in} provenance.

\subsubsection{Vega-Lite Parsing}
Vega-Lite is a high-level grammar for interactive data visualization \cite{satyanarayan2016vega}.
\gram{This grammar} supports declarative specification of unit visualization through a JSON object, which can be formally represented as the following four-tuple:
$$ unit := (data, transforms, mark\mbox{-}type, encodings) $$
In {\NLI} systems, the four-tuple reflects the query interpretation results of three aspects.
Therefore, we parse the JSON object to reveal the visualization process.

According to the execution logic of the grammar, the initial dataset with attributes is first loaded from the location declared in the \textit{data} field. 
Then, transformation tasks (if specified) are performed in the order specified in the \textit{transformation} field. 
Aggregation tasks (e.g., sum, mean) are \gram{sometimes} specified in the \textit{encoding} fields, so we extract them from the visual encoding. 
To do \gram{so}, we first identify the encoded attributes without an \textit{aggregation} field or with a \textit{bin} field, and use these attributes as dimensions to group data records, then aggregate the remaining attributes to \gram{obtain} the derived values.
To visualize the transformed data, we extract the chart type in the \textit{mark} field and the encoding scheme in the \textit{encoding} field to generate a new version of the Vega-Lite specification.
This new version of visualization keeps the connection with the transformed data and supports visual cues (e.g., highlighting) and user interactions. 

\subsubsection{Data Selection}
Inspired by the prior work \cite{narechania2021diy}, we add smart constraints to choose data subsets with relevant columns to achieve an efficient presentation. 
For the filtering task in \autoref{fig:approach}$B$, smart constraints require the data subset to contain both data records that satisfy and do not satisfy the filtering condition.
As a result, some data records in the data subset are filtered out while others are retained, making the task visible by comparing the data tables before and after the task.

The attributes specified in the Vega-Lite specification are chosen as relevant columns, allowing users to focus on useful information.
However, presenting only the relevant columns is not intuitive for users to trace the data items between tasks. 
The system chooses one attribute as a key attribute and presents it throughout the sequence of data provenance. 
The key attribute needs to meet two requirements: 
\begin{enumerate}
\item Uniquely identifies the data item, distinguishing itself from other data items.
\item Provides semantic information \gram{that is easy to understand} ({\eg} each person's name).
\end{enumerate}
Therefore, we \gram{select the} nominal or ordinal attribute with the lowest repetition rate \gram{as the key attribute}. 
In \autoref{fig:approach}$B$, the Title field is used as the key attribute of the IMDb movie dataset to identify each data item.

\subsubsection{Visual Cues}
The system leverages visual cues to emphasize the changes in sample data provenance.
For filtering tasks, the table columns of relevant attributes are highlighted in orange, and the filtered data items are displayed in translucent gray with a strike-through line. For aggregation and distribution tasks that group the data records by an attribute to calculate the new derived values, the same values of the grouped attribute are merged into the same cell.
Visual cues are also used to connect \gram{the} data tables and the visualization. The corresponding record in the data table will be highlighted when hovering on an element in the visualization charts.

\subsection{Hint Generator}
Hints are designed to help users in query phrasing and revision. 
Based on the results analysis of the pilot study (Section \ref{sec:pilot}) and survey of prior work, we design \gram{the} Hint Generator to \gram{produce} hints during the data analysis.
The system integrates a set of well-known rules to detect errors regarding attributes, tasks and visual encodings, and provides hints alongside the returned charts.
For the problems from unexpected query interpretation, the system infers the potential errors according to visualization adjustments \gram{through} widgets and provides hints after user interactions.

\subsubsection{Ruled-based Hints}
Rule-based hints are geared towards the problems in users' queries. 
The system first examines the query sentences and alerts users to the unused but potentially important keywords ({\eg} \textit{at least}).
Then, the system checks the rest of the query sentence from the following aspects.

For \textbf{attributes}, misspellings and ambiguities are common during the input.
The word in the query \gram{is} identified as a misspelling if it is ``close'' but not equal to an attribute name (measured by Levenshtein distance \cite{levenshtein1966binary}).
We set the threshold of ``close'' to $ 0.2 * len(attributeName)$ \gram{based on} prior work \cite{yu2019flowsense}. 
The ambiguities can be detected when multiple attributes are inferred from the same word.

For \textbf{tasks}, sometimes users specify \textit{impractical parameters} \gram{because of} a lack of knowledge of the dataset. 
For example, an out-of-range filter criteria results in an empty chart with no data item.
Our system records the changes in the data column during the provenance and informs users when the dataset becomes empty in the process.

For \textbf{visual encodings}, the system integrates a set of well-known rules to check the encoding scheme, including the chart type and encoding channel.
When users pose unsuitable schemes for visual encoding ({\eg} specifying a bar chart for two quantitative attributes), the system \gram{generates} the requested chart type ({\eg} bar chart) and \gram{provides} a hint for \gram{the} recommended chart type ({\eg} scatterplot).

\subsubsection{Interaction-based Hints}
Since the interactive widgets enable users to correct errors, the system provides hints as feedback from user interactions to help them gain knowledge of query phrasing.
The hints are generated based on the analysis of interaction logs and the original queries and consist of two major components, including target suggestions for local revision to the original query and valid query examples that can be used for subsequent analysis.
The generation of query examples \gram{is} introduced in Section \ref{sec:example}.

For \textbf{attributes}, when a new attribute is added to the process and is not extracted before, the system considers this a \textit{failure of inference} and suggests users explicitly mention the attribute in the query. 
If the attribute is extracted but used for other tasks, which means the {\NLI} constructs an unexpected parsing dependency of sentence structure, the system suggests users rephrase the sentence and provide query examples as alternatives.
When unexpected attributes are implicitly extracted and removed by users, the system infers the error of \textit{unexpected types} and gives hints to users to avoid the corresponding words in the original query. 

For \textbf{tasks}, multiple alternatives \gram{may be available} when it is underspecified in the query.
For instance, the {\NLI} chooses \textit{Mean} as the default aggregation type. 
If users switch to other types ({\eg} \textit{Sum}), the system considers this \textit{unexpected type} of task and suggests adding related words ({\eg} \textit{total}) to the query for explicit specification.
Filtering tasks are usually partially correct in terms of operators ({\eg} ``$>$'') and parameters ({\eg} \textit{100,000}), and users can easily modify the unexpected parts.
After user correction, the system suggests supported phrases ({\eg} \textit{more} and \textit{over}) to revise the related parts of the queries.

For \textbf{visual encodings}, users often modify the unexpected \textit{chart types} using interactive widgets, which is intuitive and efficient for \gram{obtaining} the desired results.
The system \gram{suggests} an explicit specification of the desired chart type in the query when it is not the default option for the \gram{analytical} intents.
Since \textit{encoding channels} cannot be specified through NL queries in NL4DV, the system does not provide hints for the exchange of encoding channels.

\subsubsection{Query Example Generation}
\label{sec:example}
After user adjustments through widgets, the system considers the final visualizations as the desired results and generates query examples to help the users in query phrasing.
The query examples are generated by a template-based method, which has been shown to be effective and controllable in specifying known analysis intents \cite{srinivasan2021snowy}.

Our method takes the Vega-Lite object of the final result as input, extracts the visualization process from \textit{transform}, \textit{mark}, and \textit{encodings} fields, and translates \gram{them} into linguistic expression.
We prepare a set of query templates with placeholders based on the utterances collected in the prior {\NLI} study \cite{srinivasan2021collecting, fu2020quda}.
For example, multiple templates are available for a scatterplot with two quantitative attributes.
\begin{itemize}
    \item \textit{Show \underline{attribute1} and \underline{attribute2} in scatter plot.}
    \item \textit{What is the relationship between \underline{attribute1} and \underline{attribute2}?}
    \item \textit{How are \underline{attribute1} and \underline{attribute2} correlated?}
\end{itemize}
The query templates can be further extended by phrases ({\eg} \textit{for Action movies}) and clauses ({\eg} \textit{whose release year is after 2002}) to specify additional filtering tasks.
A generated query example is shown in \autoref{fig:approach}$C$.
Such translation ensures a complete expression of the visualization results and can be easily understood by the users.

To assess the feasibility of the generated examples, we employ NL4DV as a discriminator to parse the example queries and compare their results with the current result adjusted by the users.
Finally, we randomly recommend one of the valid examples to the users.
\begin{figure*}[!t]
  \centering
  \includegraphics[width=\linewidth]{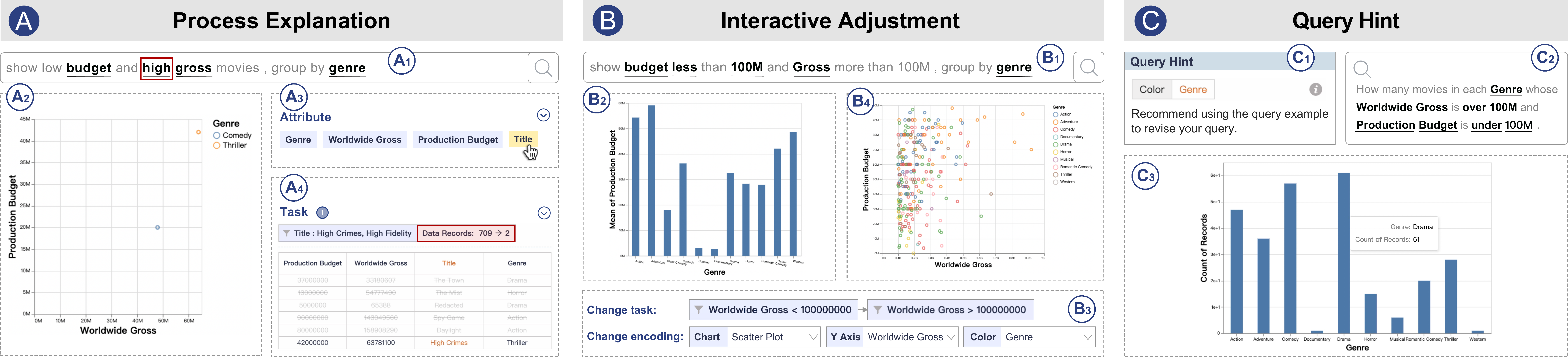}
  \caption{The first scenario illustrates how the system helps the user increase his knowledge of NLI capability. (A) From the process explanations, the user identifies an unexpected filtering task and learns that the NLI can not understand vague modifiers. (B) He tries a new query and corrects errors through interactive adjustments. (C) He leverages the query example in the hint to formulate a new query and verify the data insight. Through the query adjustments, he learns how to better phrase the queries.}
  \label{fig:case_one}
  \vspace{-5pt}
\end{figure*}

\section{Usage Scenario}
\begin{figure*}[!t]
 \centering 
 \includegraphics[width=\linewidth]{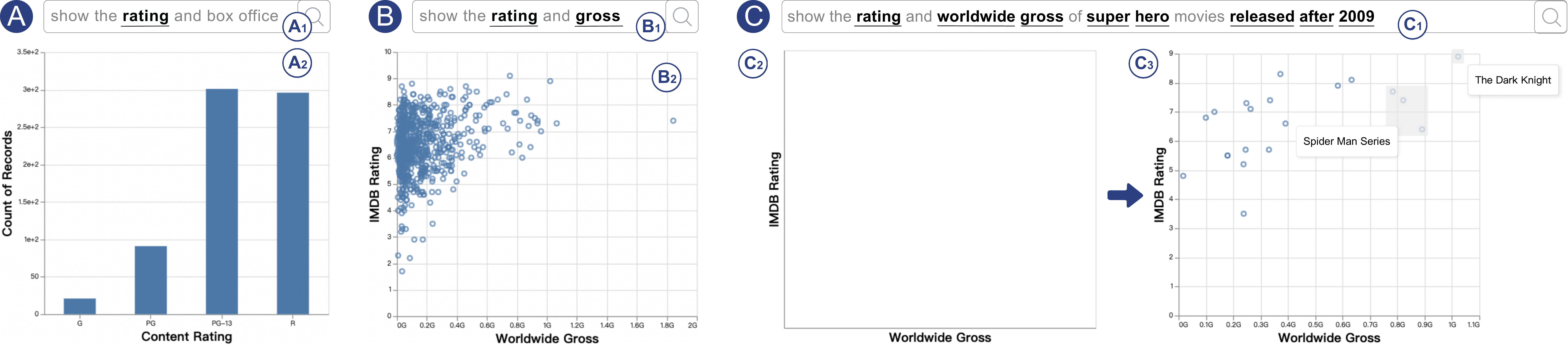}
 \caption{The second scenario presents how {\name} helps the user increase the knowledge about the dataset to be explored.} 
 \label{fig:case_two}
 \vspace{-5pt}
\end{figure*}

This section presents two \gram{usage scenarios}. The first \gram{scenario} (\autoref{fig:case_one}) presents how the system helps users increase their knowledge about the {\NLI} capability. 
The user inputs queries that {\NLI} cannot support well, and the system helps him revise the query formulation. 
The second \gram{scenario} (Figure \ref{fig:case_two}) shows how the system helps users increase the knowledge about the data to be explored. 
In this \gram{scenario}, the user may pose queries that contain inaccurate expressions of attributes and wrong parameters of tasks. The system helps him locate and correct the unexpected results.

\subsection{Scenario 1: Increase the Knowledge of {\NLI}}
In this scenario, the analyst is a novice user of {\NLI} tool and wants to analyze the IMDb movies and gain insights \gram{to inform} movie investment. 
He first loads the movie dataset and spends some time familiarizing himself with this dataset.
Being interested in the cost-effective movie genres, he inputs the query \textit{``show low budget and high gross movies, group by genre''} (\autoref{fig:case_one}$A_1$). 
In response, the system generates a scatter plot to visualize the correlation between Production Budget, Worldwide Gross, and Genre (\autoref{fig:case_one}$A_2$).
However, the chart has only two points.
The analyst feels confused about the result, so he turns to the Explanation panel to understand the visualization process.
He finds an attribute Title with uncertainty in the process (\autoref{fig:case_one}$A_3$), and it is inferred from the keyword ``\textit{high}'' in the query (\autoref{fig:case_one}$A_1$), which means that ``\textit{high}'' is treated as a specific value of movie Title instead of the data range of Worldwide Gross.
As a result, only two movies containing ``high'' in the Title are left (\autoref{fig:case_one}$A_4$) and visualized in the resulting chart.
 
With the help of the explanations, the analyst realizes that the {\NLI} can not understand vague modifiers of the data range, so he \feng{clarifies the concepts of ``high'' and ``low'' with specific data ranges}: \textit{``show budget less than 100M and Gross more than 100M, group by genre''} (\autoref{fig:case_one}$B_1$). 
The system generates a bar chart to show the average Production Budget across different Genres (Figure \ref{fig:case_one}$B_2$). 
Besides the unexpected chart type, the analyst also notices an incorrect filtering task ``Worldwide Gross $<$ 100,000,000'' from the Explanation panel.
However, the analyst fails to get the desired result through query revision.
So he turns to using interactive widgets to modify the process (Figure \ref{fig:case_one}$B_3$), including modifying the filtering tasks and switching the chart type to scatterplot.
From the adjusted chart (Figure \ref{fig:case_one}$B_4$), the analyst observes that the Drama genre seems to have the most movie items with high gross and low budget.

To verify this finding, the analyst leverages the query example \gram{on} the Hint panel (\autoref{fig:case_one}$C_1$) and makes simple edits to \textit{``How many movies in each Genre whose Worldwide Gross is over 100M and Production Budget is under 100M''} (\autoref{fig:case_one}$C_2$).
\feng{The system interprets the query and returns a bar chart (\autoref{fig:case_one}$C_3$)} showing that the Drama genre has the largest number of movies that meet the filter criteria.
From this, the analyst also learns that such sentence structures are more likely to be correctly parsed by the {\NLI}.

\subsection{Scenario 2: Increase the Knowledge of Data}
In this scenario, the analyst is not entirely familiar with the dataset.
Being intrigued by the correlation between movie rating and revenue, he inputs \textit{``show the rating and box office''} (\autoref{fig:case_two}$A_1$) and expects a scatterplot.
However, the system renders a bar chart that only presents the data distribution of Content Rating (\autoref{fig:case_two}$A_2$). 
The analyst considers it an unexpected result and tries to fix the problems. 

From the query input box (\autoref{fig:case_two}$A_1$), he finds that the keyword \textit{``box office''} is not identified. 
The analyst realizes that {\NLI} fails to infer the attribute Worldwide Gross, so he revises the query with the correct attribute name (\autoref{fig:case_two}$B_1$).
Meanwhile, a hint for \textit{``Content Rating''} reminds him to resolve the ambiguity of \textit{``rating''} in the NL query. 
He chooses \textit{``IMDB Rating''} through an interactive widget and receives a new scatterplot (\autoref{fig:case_two}$B_2$).
From the chart, the analyst does not find the apparent correlation between gross and ratings, so he wants to dive into some movies from creative types of interests and released in later years.
He modifies the query to \textit{``show the rating and worldwide gross of super hero movies released after 2009''} (\autoref{fig:case_two}$C_1$). 

The system has learned user preference for \textit{``rating''} and takes IMDB Rating by default (\autoref{fig:case_two}$C_2$).
However, the chart is empty with no data point.
Fortunately, the Hint panel helps the analyst to address the problem: \textit{``No records satisfy the filter criteria.''}
The Task aspect of the Explanation panel indicates that only 20 Super Hero movies \gram{are} in the dataset, and none of them are released after 2009.
Therefore, the analyst deletes the filtering task for \textit{``Release Year''} and gets the non-empty scatterplot in return (\autoref{fig:case_two}$C_3$).
Based on the chart analysis, the analyst finds that the movie with the highest Worldwide Gross ({\ie} The Dark \gram{Knight}) gets the highest IMDB rating.
\gram{By} contrast, in the Spider-Man series, the movies with higher gross \gram{have} lower ratings.
\section{Evaluation}
We conduct a user study to evaluate the system's effectiveness and usability in explaining and diagnosing {\NLI}-based data analysis. 
We also examine the impact of the explanations on the analysis process, including whether the explanations increase the knowledge about the {\NLI}, disrupt the analysis process, and increase the trust in {\NLI}-based data analysis.

\subsection{Participants and Setup}
We recruited 12 participants (P1-P12, three females, nine males, aged 24-32) from a technology company's business intelligence department and the national laboratory of a local university.
Four participants (P1-P4) were senior data analysts with more than five years of working experience and proficiency in programming languages ({\eg} Python, SQL) and data analysis software ({\eg} Tableau \cite{tory2019mean}, Microsoft Power BI \cite{powerbi}).
Four participants (P5-P8) were data analysis researchers with over four years of visualization experience and have published related research papers.
The remaining four participants (P9-P12) were graduate students at a local university. They rated themselves as intermediate users or beginners when we asked them about their expertise in data analysis and {\NLI} tools. 
Each participant was compensated with \$15 in cash after completing our studies.

To conduct a comparative study, we used the system in the pilot study as the baseline which is the ablated version of {\name} without the explanation components.
Both systems share the same functions to support dataset selection (\autoref{fig:system}$A_1$), data overview (\autoref{fig:system}$A_2$), and NL-based data analysis (\autoref{fig:system}$B_1$,$B_2$). The baseline does not provide users with the Explanation panel about the {\NLI} process and the query input box (\autoref{fig:system}$B_1$) does not provide visual cues to highlight keywords in the query.

\subsection{Procedure and Tasks} 
The whole pilot study consisted of four stages as follows and lasted for about 2 hours.
We introduce two types of tasks ({\ie} target replication and open-ended exploration) for different analysis scenarios, which have been widely used in prior works \cite{srinivasan2020ask, gao2015datatone, setlur2016eviza, hoque2017applying, srinivasan2020inchorus}.
Target replication leverages the visualization targets as the ground truth of user expectations and helps us to quantify the helpfulness of our system in process understanding and problem solving.
Open-ended exploration mimics the real-world analysis process in which users perform multi-step exploration to find data insights. It helps evaluate the usability of our system and the impact of explanations on the analysis process.

\textbf{Introduction and Training (30 min).} We first briefed participants about our study. We collected their demographic information (e.g., name, sex, and age) and sought their consent to record the system interactions and the conversation for further analysis. 
Then, we provided a tutorial for each system with several analysis cases on an example dataset. 
After the demonstration, participants could familiarize themselves with the systems using the example dataset.

{\bf Target Replication (30 min).}
Participants were required to use our system and baseline to reproduce 12 visualization targets for the 2 given datasets (3 tasks for each system and dataset). In addition to the \textit{IMDb movie} dataset in the pilot study, we also included a \textit{car} dataset \gram{that contained} 304 data items with eight attributes ({\eg} Cylinders, Displacement, and Origin). We used different tasks for two systems to mitigate the learning effect on the tasks. The difficulty of the tasks in each system was balanced according to the number of attributes involved. The order of \gram{the} systems and datasets was counterbalanced.

{\bf Open-Ended Exploration (30 min). }
Participants were required to use our system and the baseline to explore the \textit{housing} dataset to provide advice for the clients to purchase homes. The \textit{housing} dataset contains 1,460 data items with 15 attributes ({\eg} Lot area, Home type, Year, and Price). The advice should be supported by the data insights found in the exploration. Participants were encouraged to explore the data following a realistic and natural workflow. We did not restrict the number of attempts or the details of exploration.

\begin{figure*}[tb]
 \centering 
 \includegraphics[width=\textwidth]{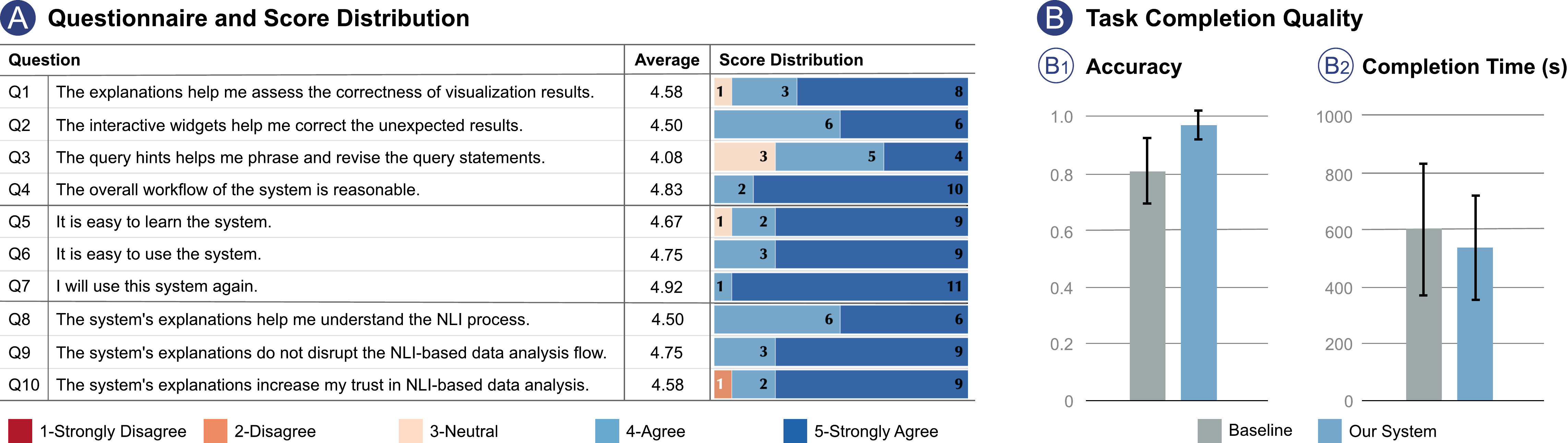}
 \caption{The results of user interviews and task completion analysis. (A) The questions in the questionnaire, together with the average values and the distribution of the ratings. (B) The task completion quality in terms of accuracy and completion time.}
 \label{fig:evaluation}
 \vspace{-5pt}
\end{figure*}

\textbf{Semi-structured Interview (30 min).}
After the exploration tasks, \gram{the} participants needed to complete a questionnaire with 10 questions.
We provided five-point Likert-scale questions that required participants to express their opinions ranging from ``strongly disagree'' (1) to ``strongly agree'' (5). 
The questions are listed in Figure \ref{fig:evaluation}$A$.
Then, we conducted a semi-structured interview to collect user feedback regarding their exploration experience using NLIs.

\subsection{Results and Analysis} 
\gram{Considering} the quantitative analysis of the target replication tasks, \feng{the user behaviors in response to system results,} the ratings of the questionnaire, and the user feedback from the semi-structured interviews, we discuss the effectiveness and usability of our system, as well as the impact of the explanations on the NLI-based analysis process.

\subsubsection{Task Completion Quality}
For the target replication tasks, we conducted a quantitative comparison between our system and the baseline in terms of task accuracy and completion time.
In \autoref{fig:evaluation}$B_1$, we calculated the average task \textbf{accuracy} \feng{(the proportion of the completed tasks)} at the individual level with 95\% confidence interval.
On average, our system ($ \mu $ = 0.96, $ \sigma $ = 0.07) helped users to achieve a higher accuracy compared with the baseline system ($ \mu $ = 0.81, $ \sigma $ = 0.11). 
The Wilcoxon signed-rank test shows a significant difference (p = 0.0066) in average accuracy between the two systems.
For task \textbf{completion time} in \autoref{fig:evaluation}$B_2$, it generally took less time for participants to complete the tasks using our system ($ \mu $ = 540.08, $ \sigma $ = 174.42) than using the baseline ($ \mu $ = 599.75, $ \sigma $ = 222.21).
However, no significant difference (p = 0.2036) \gram{was observed} between the two systems. 
This result indicates that the explanations of our system did not hinder task efficiency.


\feng{
\subsubsection{User Behaviors}
Together with the participants, we reviewed the recordings of system operations and oral comments during the tasks to analyze the use of explanations. We identified three kinds of user behaviors in response to system results.
\begin{itemize}
    \item \textbf{Use (view only):} Participants viewed the process explanations only and revised the queries on their own if they identified problems.
    \item \textbf{Use (view \gram{and} adjust):} Participants viewed the process explanations and adjusted the process with the help of interactive widgets or query hints.
    \item \textbf{Not use:} Participants neither viewed the process explanations nor adjusted them.
\end{itemize}
As shown in \autoref{tab:behavior}, the participants preferred using the explanation components to diagnose the process and fix the problems.
The process explanations were frequently used (80+45=125 out of 149) to confirm the expected system results, and sometimes (45 out of 149) even supported interactive adjustments for follow-up exploration.
When encountering unexpected system results, participants relied on the process explanations to locate the problems (37+42=79 out of 95) and adopted different correction strategies depending on the \gram{complexity of the} problems.
Specifically, they preferred using interactive widgets if the problem required a minor change ({\eg} only need to modify the parameter of a filtering task), while they tended to revise the query on their own when the problem required multi-step modification operations.
However, they might switch to interactive widgets after several failed attempts at query revision.
Query hints \gram{were} very helpful in case of empty results (caused by empty data records or unsupported chart types).
Sometimes participants were confident in the system results or tried to locate and correct the problems on their own, so they did not use the explanation components (16 out of 95).
}

\begin{table}[t]
\setlength{\abovecaptionskip}{0cm}
\caption{\feng{Distribution of participants' use of explanation components \\ in response to system's parsing results.}}
\label{tab:behavior}
\resizebox{\linewidth}{!}{
\renewcommand{\arraystretch}{1.5}
\begin{tabular}{l|llll}
\hline
\textbf{Parsing results} & \textbf{Use (only view)} & \textbf{Use (view and adjust)} & \textbf{Not use} & \textbf{Total} \\ \hline
\textbf{Expected} & 80 & 45 & 24 & 149  \\ \hline
\textbf{Unexpected} & 37 & 42 & 16 & 95  \\ \hline
\textbf{Total}     & 117 & 87 & 40 & 244  \\ \hline
\end{tabular}}
\vspace{-5pt}
\end{table}

\subsubsection{Effectiveness}
According to the questionnaire and interviews, most participants (8/12 strongly agree, 3/12 agree) agreed that the \textbf{process explanations} helped them assess the correctness of the visualization results (Q1). 
P5 said \textit{``There is no doubt that the explanations can help me to identify problems in the process, especially for some tasks `behind' the visualizations.''}
\feng{P10 mentioned that the data provenance is intuitive and helps her understand how the tasks operate on the data, such as the distribution and trend tasks.}
Some participants appreciated the visualization of the query-visualization connections and the inference uncertainty for problem location.
P7 commented \textit{``Once I get the explanations of the process, I always check the uncertainty in the process first and compare it with the highlighted content in the query. It really helps me improve the efficiency of problem checking.''}

Participants also provided positive feedback (6/12 strongly agree, 6/12 agree) to the \textbf{interactive widgets}, praising its ability to support flexible adjustments to get the desired results (Q2), even some that could not be obtained through natural language queries.
P4 mentioned \textit{``I particularly like using interactive widgets to change the visual encoding of results, as I find it impossible to specify a specific visual encoding channel for an attribute through natural language.''}
P3 suggested extending interactive widgets to support more complex scenarios, \textit{``It would be better to support the set operations for multiple tasks, such as union (`or' operation).''}

Most participants thought that the \textbf{query hints} helped them phrase and revise the queries (Q3), especially helping users \textit{``avoid unreasonable errors''} (P6) and \textit{``improve the efficiency of query modification''} (P5).
Participants were eager to use the query examples for the following exploration.
\feng{P3 and P5 mentioned that they preferred leveraging the suggested query examples for new analysis rather than entering the queries from scratch, which improved the input efficiency and accuracy.}
P9 said \textit{``I can just modify the content of the query example while keeping the syntactic structure. Such queries are more likely to be parsed correctly.''}
We also discussed with the participants the quality of the query examples.
Participants thought that the provided query examples were \textit{``easy to understand''} (P11) and \textit{``easy to follow''} (P7).
Regarding the naturalness of the query examples, P8 commented \textit{``The query examples are clear in expression, but sometimes a little wordy.''}

The \textbf{overall workflow} of the system is considered reasonable (10/12 strongly agree, 2/12 agree) according to user feedback (Q4).
P8 specified \textit{``The process explanations help me to identify unexpected results, so I can fix simple problems by myself and solve more complex ones with the help of widgets and hints. The system workflow is very fluent.''}
The correctness of data analysis is highly valued in the industry. P1 pointed out \textit{``The workflow improves the rigor of {\NLI} analysis and will facilitate the application of {\NLI} analysis in industrial analysis scenarios.''}

\subsubsection{Usability}
The system usability centered on user experience regarding learning cost and workload.
Most participants (except P9) agreed that the system is \textbf{easy to learn} (Q5) and \textit{``intuitive''} (P1, P5, P10).
P1 added that \textit{``I feel familiar with the system because it reminds me of Jupyter Notebook, where I can debug the codes block by block based on the intermediate results and system logs, such as errors and warnings.''}
However, P9 (with a rating of 3) posed his concerns about the learning cost because \textit{``the system introduces a series of features that took me some time to fully understand.''}
Nevertheless, all participants (9/12 strongly agree, 3/12 agree) agreed that the system is \textbf{easy to use} (Q6) and most of them mentioned that the system design is not complex.
P9 reported that \textit{``the system is very user-friendly, logical, and beautifully designed. I feel very comfortable using this system.''}
Finally, all participants (11/12 strongly agree, 1/12 agree) are willing to \textbf{use our system again} in the future.

\subsubsection{Impact on Visual Analysis}
We further investigate the impact of explanations on the {\NLI}-based visual analysis in terms of understanding, fluidity, and trust.
According to participants' rating (Q8) and feedback, the explanations help \textbf{understand the {\NLI} process} (6/12 strongly agree, 6/12 agree) in two main ways.
First, our system improves the users' knowledge about the {\NLI} pipeline.
P10 stated \textit{``I realized that (inference of) attributes, tasks, and visual encodings are important steps for {\NLI} to process queries, so I organize my query statements according to these perspectives.''}
Second, our system reveals the interpretation capability of {\NLI}, including the supported tasks ({\eg} correlation) and the corresponding triggered keywords ({\eg} \textit{correlate}, \textit{relationship}). 
P4 commented \textit{``I think the {\NLI} is not perfect, and the explanations helped me to identify its strengths and weaknesses, so I can use it better for my analysis.''}

As an introduced feature, the explanations (Q9) \textbf{do not disrupt {\NLI}-based data analysis} (9/12 strongly agree, 3/12 agree).
The results analysis of the target replication task shows that our system improves the efficiency of task completion compared to the baseline system.
In the interview, P7 mentioned that \textit{``the functions of explanations are integrated into the same area and can be turned off, so I can use only the {\NLI} part of the system when I am confident in the analysis.''}
\gram{In contrast}, the explanations improved the efficiency of the analysis.
P3 reported that \textit{``it would have taken me a lot of time to understand the query output without explanations. The system used a storytelling-like approach to explain the visualization output, which significantly accelerates my analysis progress.''}
In addition, he mentioned that the explanations also give users heuristic guidance for subsequent exploration.

Finally, most participants (except P4) agreed that the explanations \textbf{increased the trust} in {\NLI}-based data analysis (Q10).
The explanations \textit{``make the visualization process transparent''} (P6) and make the participants \textit{``feel confident after the confirmation of the intermediate process''} (P5).
However, P4 raises the concern of trusting {\NLI} (without explanations) for data analysis.
He pointed out that he remains suspicious of the current {\NLI}-based analysis tools because they still require human understanding and intervention.

\section{Discussion}
In this section, we first summarize the lessons that we learned from the user study, including providing a preview of query results for {\NLI} debugging and using explanations for NLI to promote system discoverability.
Then, we discuss the generalizability as well as the limitations and future work.

\subsection{Lessons Learned}
\textbf{Provide a preview of query results for NLI debugging.}
We observed some interesting patterns during the evaluation.
Some users could not determine the correctness of the results in the baseline system, so they tried to revise the query to test the system response. 
For example, one user changed the range of filter criteria in the query to test whether the filtering task worked in the process.
However, he \gram{was still not} sure whether the filtering task was correct. 
When receiving unexpected results, some users had to simplify the query sentences by removing some words to locate the problems in the queries.
These debugging strategies are similar to ``black-box testing'' and cost a considerable amount of time during the analysis process.
\gram{This finding} also motivates us to further study how to help users quickly preview and compare the results of different revision strategies.

\textbf{Use explanations for {\NLI} to promote system discoverability.}
The explanations increase the {\NLI} discoverability (i.e., the users' awareness and understanding of system-supported commands). 
In contrast to much prior research~\cite{srinivasan2019discovering, srinivasan2021snowy, wang2022interactive} that recommends supported utterances for next-step data analysis, the explanations reveal the {\NLI} process, enabling users to explore more alternative expressions that adapt to users' preferences and can be understood by the system.
For example, one user in our evaluation was accustomed to using parentheses or commas in specifying multi-step data operations (e.g., filters).
From the system explanations, he verified that such syntax can be well parsed by {\NLI} and preferred to use this syntax in the subsequent queries.

\subsection{Generalizability}
Our work is based on NL4DV, a typical toolkit following the general pipeline of NLI \cite{shen2021towards, gao2015datatone}, and the major components of our system (i.e., Provenance Generator, interactive widgets, and Hint Generator) are independent of the specific details of NLI algorithms and can be directly applied to the existing NLIs. 
\gram{The} Provenance Generator reveals the visualization process based on the Vega-Lite specification, which is widely used by many {\NLI}s \cite{narechania2020nl4dv, wang2022interactive, srinivasan2021snowy, luo2021natural, luo2021synthesizing, srinivasan2021collecting, islam2022natural, fu2020quda}.
\gram{The} Hint Generator integrates a set of well-known rules \cite{yu2019flowsense, gao2015datatone, narechania2020nl4dv} to detect problems in the query and can be extended to support more scenarios in different application domains.
Other features, such as the highlighting of the algorithm uncertainty and the connection between visualizations and NL queries are based on the query interpretation in the NLI process. They can also be used by other NLIs that follow the typical NLI pipeline. Therefore, our work is general and can be easily used for other NLIs.

\subsection{Limitations and Future Work}
\textbf{Improve the query examples.}
The query examples in query hints are generated based on the templates collected in the prior {\NLI} study \cite{srinivasan2021collecting, fu2020quda}.
Despite its effectiveness and controllability, the approach may generate long structured sentences for complex cases with multiple tasks which do not seem as natural as those produced by humans.
One of our future works is improving the quality of the generated query examples.
Recently, generative language models \cite{brown2020language} trained on a large corpus \gram{have been} proposed to facilitate various downstream tasks \cite{wu2022ai, jiang2022discovering}.
They open up the possibility of enhancing the diversity and naturalness of linguistic expression while maintaining the accurate representation of the users' intentions.

\feng{\textbf{Improve the system evaluation.}
We show the effectiveness of our system through a comparative study with a baseline NLI that only maintains the basic functions without explanation components.
However, it would be better to design and implement a stronger baseline ({\eg} NLI with simple ambiguity widgets \cite{gao2015datatone, setlur2016eviza, hoque2017applying, tory2019mean}) for comparative study.
\gram{And the comparison can} further help reflect the values of the systematic explanations for the NLI process.}

\textbf{Provide explanations for follow-up queries.}
{\name} provides explanations for one-off queries that specify the entire visualization process in a single sentence.
Several \gram{studies} \cite{setlur2016eviza, hoque2017applying, kumar2016towards, fast2018iris, mitra2022facilitating} have supported conversational data analysis, enabling users to pose follow-up queries (e.g., ``\textit{Show only movies released after 2008}'') in the context of prior queries (e.g., ``\textit{show the distribution of production budget in a bar chart''}).
It is a flexible means of expressing analytical intents but also introduces a larger risk of incorrect query interpretation \cite{hoque2017applying}.
The system should provide explanations for follow-up queries to present how the {\NLI}s leverage the context information to interpret the follow-up queries.

\textbf{Learn from users.}
The explanations accelerate the user's learning of {\NLI} by revealing the visualization process and increasing the discoverability of the system-supported queries.
The system provides interactive widgets to support further specification of user expectations, which also provide great opportunities for {\NLI} to learn from users. 
We have enabled the system to learn the user preference to resolve the ambiguities in the following queries.
\gram{A} beneficial \gram{task is exploring} how to support user explanations for their queries and then translate them into personalized features to better interpret the queries and generate desired responses.

\section{Conclusion}
This \gram{study} focuses on explaining and diagnosing {\NLI}-based visual analysis.
We first identify and summarize the problems of unexpected visualization results during the {\NLI}-based visual analysis that require process explanations.
Then we present {\name}, an explainable {\NLI} system to illustrate the generation process of visualization results. Specifically, it reveals the process from the aspects of the attributes, tasks, and visual encodings and generates sample data provenance based on the Vega-Lite specification.
To help users correct errors, the system provides interactive widgets to adjust the process and generates hints to refine the queries. 
We introduce two \gram{usage scenarios} and a user study to demonstrate the effectiveness and usability of our system in improving the task accuracy and efficiency of {\NLI}-based data analysis. 

In the future, we plan to enhance the explainability of {\name}, including revealing more details of query interpretation, supporting a higher level of control and personalization of the {\NLI} process, and providing explanations for follow-up queries in conversational visual analysis.
We \gram{also aim} to improve the quality of the generated query examples to better assist users in query revision.
Moreover, we plan to conduct longitudinal studies to further investigate the impact of systematic explanations in building trust in {\NLI} results and facilitating human-machine collaboration.

\section*{Acknowledgment}
We would like to thank Wei Zhang and Jiacheng Pan for their kind help. We also thank the anonymous reviewers for their insightful comments. This paper is supported by National Natural Science Foundation of China (62132017) and Fundamental Research Funds for the Central Universities (226-2022-00235).


\ifCLASSOPTIONcaptionsoff
  \newpage
\fi



\bibliographystyle{IEEEtran}
\bibliography{main}

%





\begin{IEEEbiography}[{\includegraphics[width=1in,height=1.25in,clip,keepaspectratio]{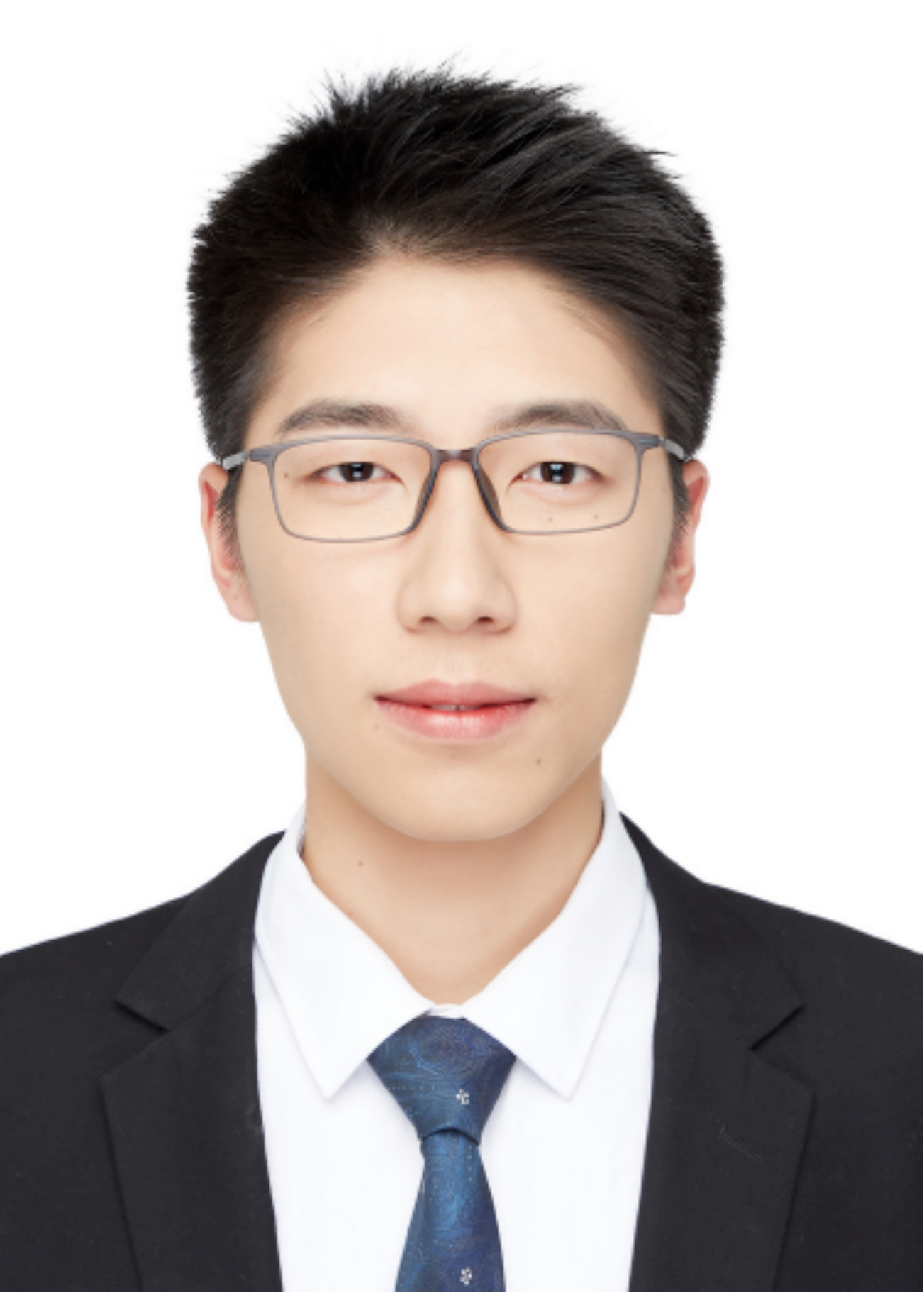}}]{Yingchaojie Feng}
is currently a Ph.D. candidate in the College of Computer Science and Technology, the Zhejiang University, China. He received the B.E. degree in software engineering from the Zhejiang University of Technology, China in 2020. His research interests include natural language processing and visual analysis.
\end{IEEEbiography}

\begin{IEEEbiography}[{\includegraphics[width=1in,height=1.25in,clip,keepaspectratio]{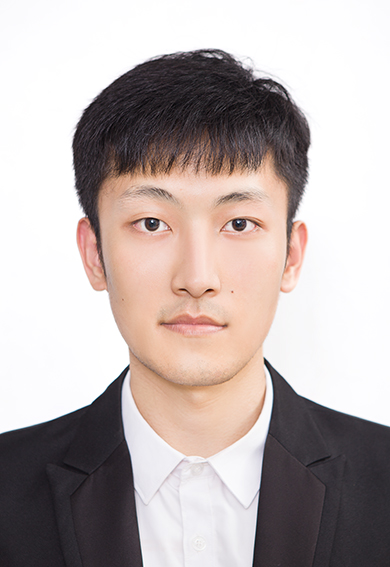}}]{Xingbo Wang}
is currently a postdoctoral fellow in the Department of Computer Science and Engineering at the Hong Kong University of Science and Technology (HKUST), where he also received his Ph.D. degree in 2022. He obtained a B.E. degree from Wuhan University, China in 2018. His research interests include human-computer interaction (HCI), data visualization, natural language processing (NLP), and multimodal analysis. For more details, please refer to https://andy-xingbowang.com/.
\end{IEEEbiography}

\begin{IEEEbiography}[{\includegraphics[width=1in,height=1.25in,clip,keepaspectratio]{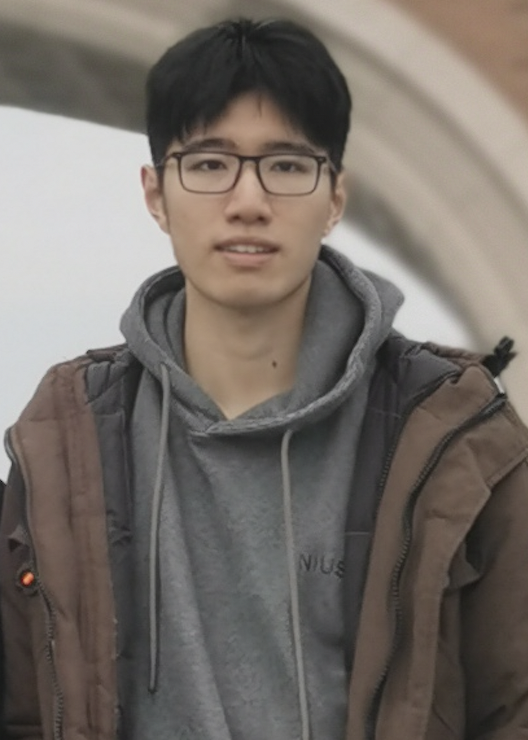}}]{Bo Pan}
received the BS degree in Electrical  and Computer Engineering from the University of Illinois Urbana-Champaign and Zhejiang University in 2022. He is currently working towards the Ph.D. degree at the State Key Lab of CAD\&CG, Zhejiang University. His research interests include visualization and deep learning.
\end{IEEEbiography}

\begin{IEEEbiography}[{\includegraphics[width=1in,height=1.25in,clip,keepaspectratio]{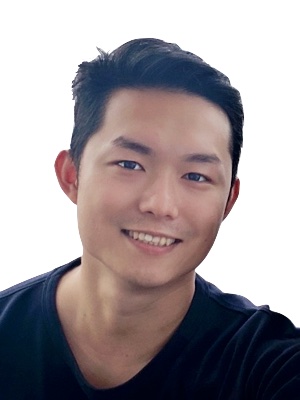}}]{Kam~Kwai~Wong} 
is a Ph.D. candidate in the Department of Computer Science and Engineering at the Hong Kong University of Science and Technology (HKUST). He received his B.E. in HKUST. His main research interests are in data visualization, visual analytics and data mining.
For more details, please refer to \url{https://jasonwong.vision}.
\end{IEEEbiography}

\begin{IEEEbiography}[{\includegraphics[width=1in,height=1.25in,clip,keepaspectratio]{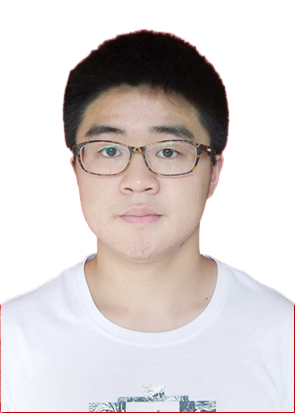}}]{Yi Ren}
received a BS degree in geographic information science from Southeast University in China in 2016. He is currently working towards the master's degree at the College of Software Engineering, the Zhejiang University. His research interests include information visualization and visual analytics
\end{IEEEbiography}

\begin{IEEEbiography}[{\includegraphics[width=1in,height=1.25in,clip,keepaspectratio]{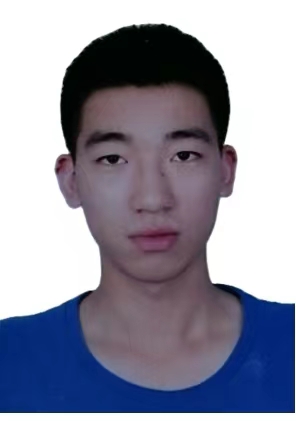}}]{Shi Liu}
received his B.S. in software engineering from the Central South University, China in 2021. He is currently a Master student in the School of Software Technology at the Zhejiang University. His research interests include information visualization, visual analysis and human-computer interaction.
\end{IEEEbiography}

\begin{IEEEbiography}[{\includegraphics[width=1in,height=1.25in,clip,keepaspectratio]{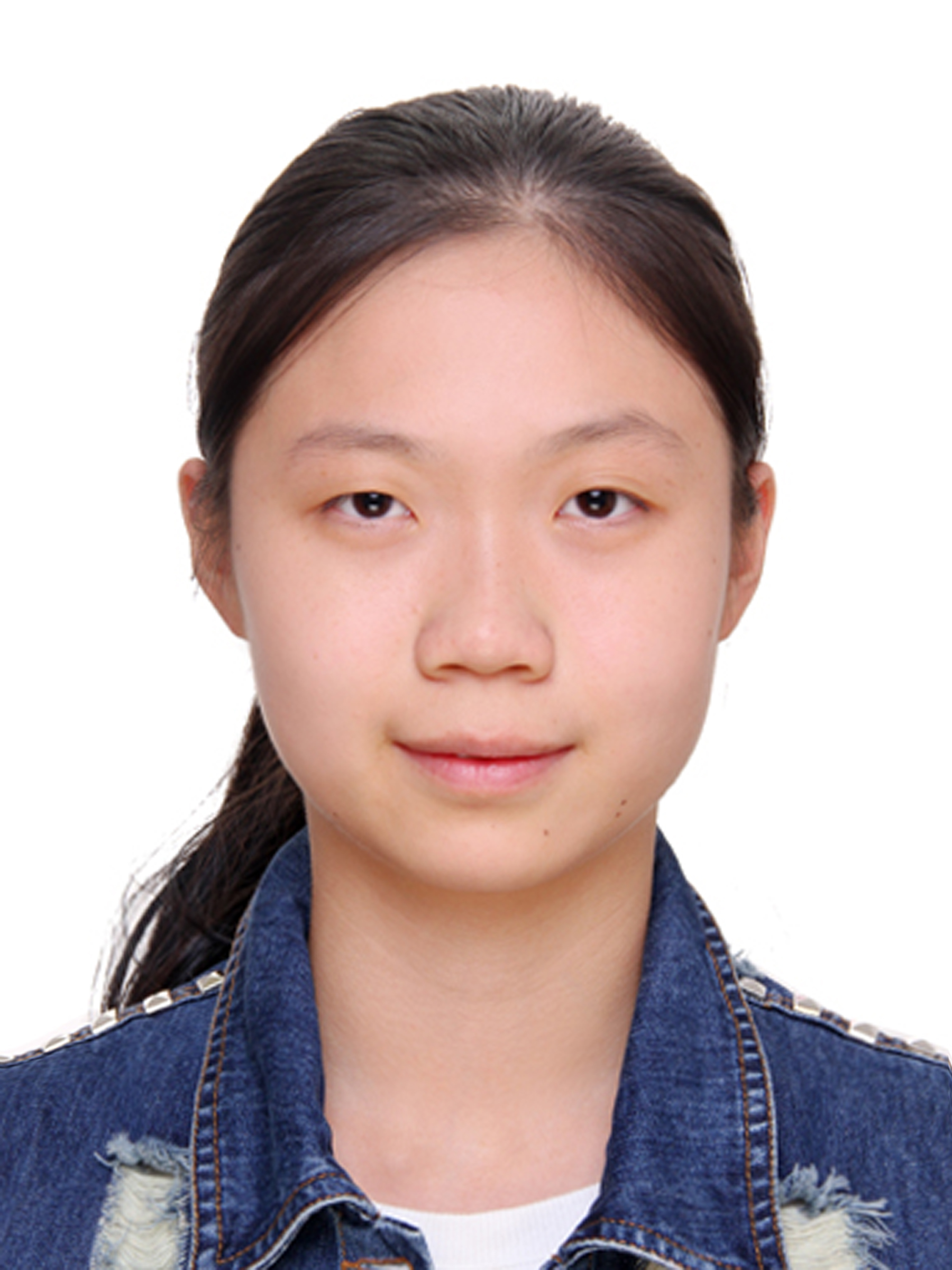}}]{Zihan Yan}
received the BS degree in computer science department from the Zhejiang University, China, in 2015. She is currently working toward the master’s degree in MIT Media Lab, United States. Her research interests include human-computer interaction (HCI), multimodal analysis and deep learning. For more details, please refer to https://yzihan.github.io/.
\end{IEEEbiography}

\begin{IEEEbiography}[{\includegraphics[width=1in,height=1.25in,clip,keepaspectratio]{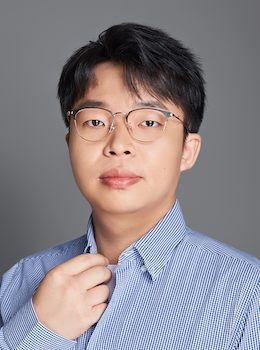}}]{Yuxin Ma}
is a tenure-track assistant professor in the Department of Computer Science and Engineering, Southern University of Science and Technology (SUSTech), China. He received his B.Eng. and Ph.D. degrees from Zhejiang University. Before joining SUSTech, he worked as a Postdoctoral Research Associate in VADER Lab, CIDSE, Arizona State University. His primary research interests are in the areas of visualization and visual analytics, focusing on explainable AI, high-dimensional data, and spatiotemporal data.
\end{IEEEbiography}

\begin{IEEEbiography}
[{\includegraphics[width=1in,height=1.25in,clip,keepaspectratio]{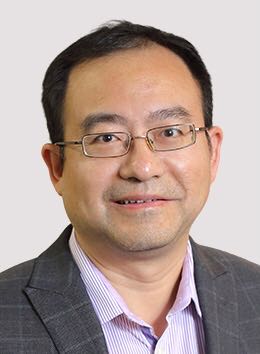}}]{Huamin Qu}
is a chair professor in the Department of Computer Science and Engineering (CSE) at the Hong Kong University of Science and Technology (HKUST) and also the director of the interdisciplinary program office (IPO) of HKUST. He obtained a BS in Mathematics from Xi’an Jiaotong University, China, an MS and a PhD in Computer Science from the Stony Brook University. His main research interests are in visualization and human-computer interaction, with focuses on urban informatics, social network analysis, E-learning, text visualization, and explainable artificial intelligence.
\end{IEEEbiography}

\begin{IEEEbiography}[{\includegraphics[width=1in,height=1.25in,clip,keepaspectratio]{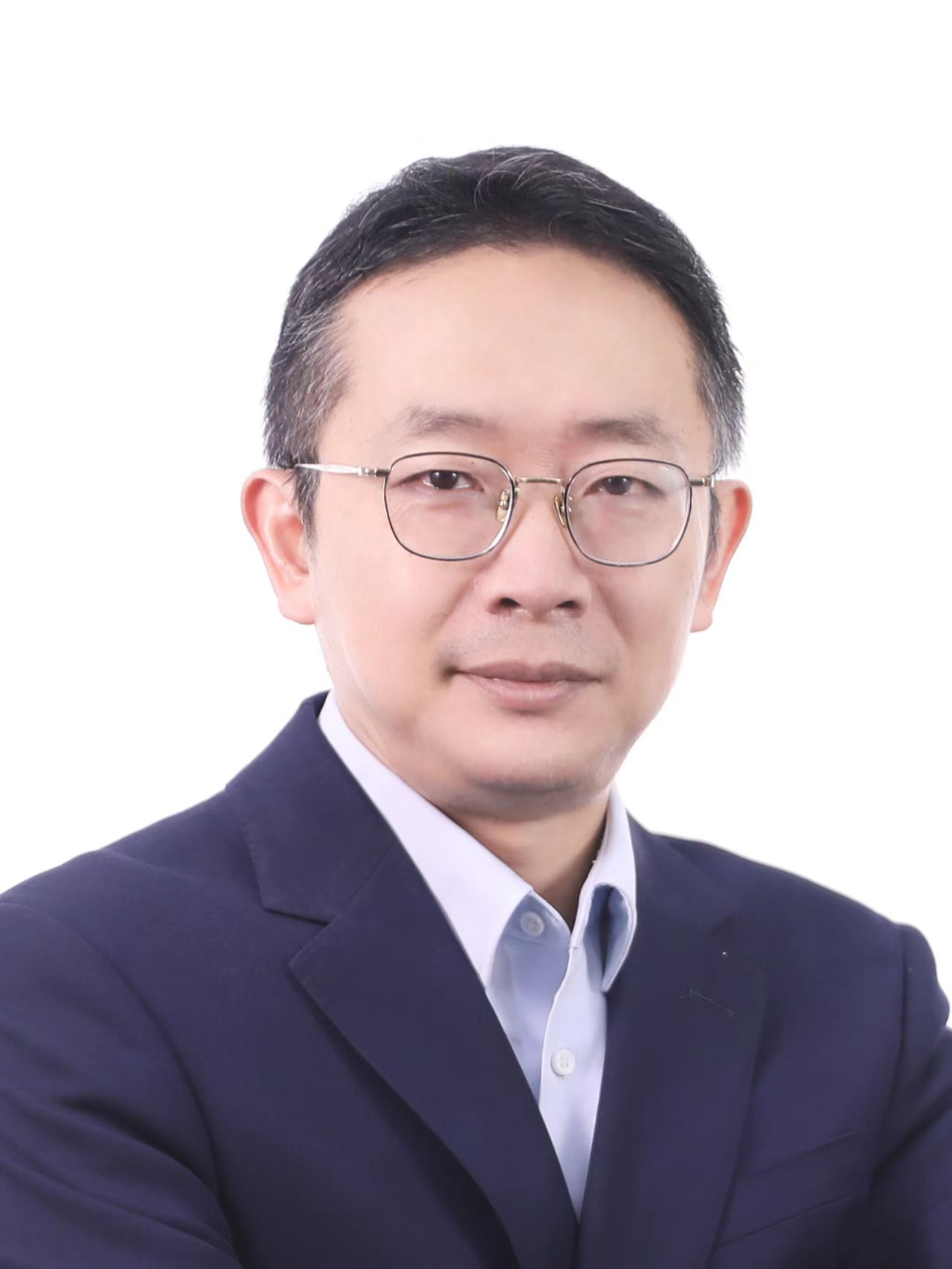}}]{Wei Chen}
is a professor in the State Key Lab of CAD\&CG at Zhejiang University. His current research interests include visualization and visual analytics. He has published more than 80 IEEE/ACM Transactions and IEEE VIS papers. He actively served in many leading conferences and journals, like IEEE PacificVIS steering committee, ChinaVIS steering committee, paper co-chairs of IEEE VIS, IEEE PacificVIS, IEEE LDAV and ACM SIGGRAPH Asia VisSym. He is an associate editor of IEEE TVCG, IEEE TBG, ACM TIST, IEEE T-SMC-S, IEEE TIV, IEEE CG\&A, FCS, and JOV. More information can be found at: http://www.cad.zju.edu.cn/home/chenwei.
\end{IEEEbiography}




\end{document}